\documentclass[%
 reprint,
 amsmath,amssymb,
 aps,
]{revtex4-1}

\usepackage{graphicx}
\usepackage{xcolor}
\graphicspath{{C:/Users/Tom/Documents/PhD/Latex Documents/Papers/Temperature Calibration and Cooling/Pictures/}}
\usepackage{grffile}
\usepackage{dcolumn}
\usepackage{bm}
\usepackage[colorinlistoftodos]{todonotes}
\begin{document}


\title{Sympathetic cooling and squeezing of two co-levitated nanoparticles}

\author{T. W.  Penny}%
 \email{Present address: Wright Laboratory, Department of Physics, Yale University, New Haven, CT}
\author{A.  Pontin}
\author{P. F. Barker}%
 \email{p.barker@ucl.ac.uk}
\affiliation{%
 Department of Physics and Astronomy, University College London, Gower St, London WC1E 6BT, United Kingdom
}%

\date{\today}

\begin{abstract}
Levitated particles are an ideal tool for measuring weak forces and investigating quantum mechanics in macroscopic objects. Arrays of two or more of these particles have been suggested for improving force sensitivity and entangling macropscopic objects. In this article, two charged, silica nanoparticles, that are coupled through their mutual Coulomb repulsion, are trapped in a Paul trap, and the individual masses and charges of both particles are characterised. We demonstrate sympathetic cooling of one nanoparticle coupled via the Coulomb interaction to the second nanoparticle to which feedback cooling is directly applied. We also implement sympathetic squeezing through a similar process showing non-thermal motional states can be transferred by the Coulomb interaction. This work establishes protocols to cool and manipulate arrays of nanoparticles for sensing and minimising the effect of optical heating in future experiments. 
\end{abstract}

\pacs{Valid PACS appear here}
\maketitle


\section{\label{sec:Intro}Introduction}

\par  The ability to cool and control the centre-of-mass (CoM) motion of levitated nanoparticles and microparticles, coupled with their extreme isolation from the environment, makes them ideal candidates for measuring weak forces.  They have been proposed as detectors in the search for dark matter candidates~\cite{Monteiro2020, Moore2021, Carney2021}, for investigating the macroscopic limits of quantum mechanics~\cite{RomeroIsart2010, BatemanWMInter2014, Bahrami2014, Weiss2021, Cosco2021} and for measuring short-range forces~\cite{GeraciZepto2016, Hebestreit2018_2, Kawasaki2020, Weiss2021, Cosco2021}. To date, only a few investigations have focused on cooling and controlling more than a single particle levitated in vacuum~\cite{Arita2018, Slezak2019}. Arrays of levitated nanoparticles are of interest as they can be used to enhance the detection of dark matter candidates~\cite{Moore2021, Carney2021}, measure vacuum friction~\cite{Zhao2012} and for evidencing the quantumness of gravity via entanglement~\cite{Sougato2017, Datta2021}. Even arrays as small as two particles can be useful for increasing the isolation from external noise sources~\cite{King1998}. A first step towards utilising arrays of levitated particles is the development of tools to control the motion of co-levitated particles.

\par Sympathetic cooling has been used extensively in ion trapping experiments to cool atomic and molecular species where no favourable internal transitions for laser cooling are available~\cite{Wineland1978, Drullinger1980, Larson1986, Baba1996, Molhave2000}. It is made possible by the coupling from the Coulomb interaction between co-trapped ions. Coupling between two levitated nanoparticles in vacuum has been demonstrated with optical binding~\cite{Arita2018} and via the Coulomb interaction~\cite{Slezak2019}. The ability to cool and control all particles via a single particle using light, while not illuminating the other co-trapped particles, can be used to minimise heating ~\cite{Millen2014, Rahman2016}, particularly when they contain internal atomic like systems such as nitrogen vacancy centers, whose internal state manipulation is highly temperature dependent.

The coupling between the particles in the array allows, in principle, to transfer other more complex motional states as well. Among these, squeezed states are considered the simplest, more easily accessible non-classical states~\cite{kimsqueezed,science_squeezing}. In the quantum regime and in presence of a multimode system, as is considered here, squeezed states are an important resource which can allow the generation and observation of entanglement between different mechanical degrees of freedom~\cite{Pontin2016,Poot2014}. Even far from the quantum domain, squeezed states have important applications for enhanced force sensing~\cite{caves_RevModPhys.52.341,parametric_amplification}. 

\par In this paper, we co-trap a pair of silica nanoparticles in a linear Paul trap that are coupled through their mutual Coulomb repulsion. By implementing a velocity damping scheme~\cite{Li2011, Tebbenjohans2018, Slezak2018, Iwasaki2019, Dania2020, Tebbenjohans2020, Kamba2021, Tebbenjohans2021, Penny2021} on just one particle we sympathetically cool the motion of the second particle to achieve sub-kelvin normal mode temperatures. This differs from previous work~\cite{ Slezak2019} were both particles were cooled simultaneously. Importantly, we also show that the Coulomb interaction can transfer other states between co-trapped particles by squeezing the normal modes of the system with a parametric drive~\cite{Rugar1991, Briant2003, Szorkovsky2011, Szorkovszky2012, Szorkovszky2013, Pontin2014, Poot2014, Pontin2016} on just one particle.

\section{\label{sec:Theory}Theory}

\par The equations of motion (EoM) for two harmonic oscillators coupled via the electrostatic force are given by

\begin{equation}
    \ddot{z}_{1} + \gamma_{1}\dot{z}_1 + \omega_{1}^2z_{1} = \frac{F_{fluct, 1}}{m_{1}} + \frac{Q_{1}Q_{2}}{m_{1}4\pi \epsilon_{0}(z_2 - z_1)^2} 
\end{equation}
\begin{equation}
    \ddot{z}_{2} + \gamma_{2}\dot{z}_2 + \omega_{2}^2z_{2} = \frac{F_{fluct, 2}}{m_{2}} - \frac{Q_{1}Q_{2}}{m_{2}4\pi \epsilon_{0}(z_2 - z_1)^2}
\end{equation}
where $z_i$ are the positions of the particles ($i=\{1,2\}$ denote the particle), $\gamma_i$ are the damping constants, $\omega_i$ are the natural frequencies, $m_i$ are the masses, $F_{fluct, i}$ are the thermal force noises defined by $\langle F_{fluct, i}(t)F_{fluct, j}(t^{\prime})\rangle = 2m_{i}\gamma_{i}k_{B}T_{0}\delta(t - t^{\prime})\delta_{i,j}$ ($j=\{1,2\}$) where $k_{B}$ is the Boltzmann constant and $T_{0}$ is the temperature of the surrounding thermal bath, $Q_i$ are the charges of the particles and $\epsilon_0$ is the permittivity of free space. Several techniques for levitating multiple nanoparticles exist but here we will focus on the case of particles co-trapped in a single linear Paul trap. Paul traps confine charged particles using a combination of static and oscillating electric fields. 

%

%

\par Considering the axial direction of a linear Paul trap where the trap is formed by only static fields, the uncoupled secular frequencies of the two particles are given by $\omega_{i} = \sqrt{\frac{2Q_{i}\kappa U_{0}}{m_{i} z_{0}^{2}}}$. Including the Coulomb interaction, the total potential for the trapped particles is given by~\cite{Morigi2001}

\begin{equation}{\label{eqn: Pot}}
    V(z_1, z_2) = \frac{1}{2}(u_{1}z_{1}^{2} + u_{2}z_{2}^{2}) + \frac{Q_{1}Q_{2}}{4\pi\epsilon_{0}|z_{2} - z_{1}|} 
\end{equation}
where $u_{i} = \frac{2Q_{i}\kappa U_{0}}{z_{0}^{2}}$. The equilibrium positions of each particle, $z_i^{eq}$, can be calculated by setting $\frac{\partial V}{\partial z_{i}} = 0$ and solving for $z_i$. From these, an equilibrium separation of

\begin{equation}\label{eqn: eqlsep}
    z_{sep}^{eq} = z_2^{eq} - z_1^{eq} = \left(\frac{Q_{1}Q_{2}(1+\frac{Q_2}{Q_1})}{u_{2}4\pi\epsilon_0}\right)^{\frac{1}{3}}
\end{equation}
can be calculated which is mass independent. Moving to a coordinate system given by the particles' deviations about their equilibrium positions, $s_i = z_i - z_i^{eq}$, and assuming $s_{i} \ll z_{sep}^{eq}$, the interaction term in Eq. \ref{eqn: Pot} can be expanded to second order about the equilibrium positions. By ignoring damping and external forces, the Euler-Lagrange EoM are found to be

\begin{gather}\label{eqn: EL}
 \begin{bmatrix} \ddot{s}_{1} \\ \ddot{s}_{2} \end{bmatrix}
 =
 -\begin{bmatrix} V_{11} & V_{12} \\ V_{21} & V_{22} \end{bmatrix}
 \begin{bmatrix} s_{1} \\ s_{2} \end{bmatrix} = -\mathbf{V} \begin{bmatrix} s_{1} \\ s_{2}\end{bmatrix}
\end{gather}
where $V_{i,j} = \frac{1}{m_{i}}\frac{\partial^2}{\partial z_{i}\partial z_{j}}V(z_{1}, z_{2})|_{z_{sep}^{eq}}$. Assuming the oscillator motion takes the from $s_{i} = s_{i,0}e^{-i\omega t}$ then the problem is reduced to finding the eigenvalues and eigenvectors of matrix $\mathbf{V}$ which describe the normal modes of the system. The eigenvalues are given by

\begin{equation}\label{eqn: efreq}
    \omega_{\pm}^{2} = \frac{\kappa U_{0}}{z_{0}^{2}}\left(A + B \mp \sqrt {C^2 + B^{2}+D}\right)
\end{equation}

where

\begin{equation}
    A = \frac{Q_1}{m_1} + \frac{Q_2}{m_2},
\end{equation}
\begin{equation}
    B = \frac{2Q_2}{(1+Q_{2}/Q_{1})}(\frac{1}{m_1} + \frac{1}{m_2}),
\end{equation}
\begin{equation}
    C = \frac{Q_1}{m_1} - \frac{Q_2}{m_2},
\end{equation}
and
\begin{equation}
    D = \frac{4Q_2}{(1+Q_{2}/Q_{1})}(\frac{Q_1}{m_1} + \frac{Q_2}{m_2})(\frac{1}{m_1} + \frac{1}{m_2}-2).
\end{equation}
Provided the eigenvalues are positive ($\omega_{\pm}^{2}>0$) then the motion is stable and the normal mode frequencies are given by $\omega_{\pm}$. The normalised eigenvectors are given by

\begin{equation}\label{eqn: evecs}
    \mathbf{e_{\pm}} = \frac{1}{\sqrt{1+r_{\pm}^{2}}} 
    \begin{bmatrix}
    1\\r_{\pm}
    \end{bmatrix}.
\end{equation}

where

\begin{equation}
    r_{\pm} = -\frac{m_{1}\omega_{\pm}^2\frac{z_{0}^{2}}{2\kappa U_{0}}(1+\frac{Q_{2}}{Q_{1}})-Q_{1}-3Q_{2}}{2Q_{2}}.
\end{equation}
%
The product $r_{+}r_{-} = -m_{1}/m_{2}$ therefore these eigenvectors are only orthogonal when $m_{1}=m_{2}$. In this case, the values of $r_{\pm}$ becomes mass independent. The eigenvectors define the normal modes in terms of the displacement of the individual particles such that

\begin{gather}\label{eqn: EL2}
\begin{bmatrix} s_{1} \\ s_{2}\end{bmatrix} = z_{+}\mathbf{e_{+}} + z_{-}\mathbf{e_{-}}
\end{gather}
where $z_{+}$ and $z_{-}$ are the amplitudes of the normal modes. For two particles with the same charge and mass (like atomic ions) the eigenvalues and eigenvectors reduce to:

\begin{equation}
    \omega_{+} = \omega_0,\,\,\,\,\,\,\,\omega_{-} = \sqrt{3}\omega_0 
\end{equation}

\begin{equation}
    \mathbf{e_{\pm}} = \frac{1}{\sqrt{2}}
    \begin{bmatrix}
    1 \\ \pm1
    \end{bmatrix}
\end{equation}
where $\omega_0 = \sqrt{\frac{2Q\kappa}{m z_{0}^{2}}U_{0}}$. In this case, we can consider $\mathbf{e_{+}}$ the in-phase CoM motion and $\mathbf{e_{-}}$ the out-of-phase stretching motion of the two particle system.

\par By assuming the same mass and size for both particles then the uncoupled EoM for the two normal modes amplitudes can be written as

\begin{equation}
    \ddot{z}_{+} + \gamma_{0}\dot{z}_+ + \omega_{+}^2z_{+} = \frac{F_{fluct, +}}{m}
\end{equation}
\begin{equation}
    \ddot{z}_{-} + \gamma_{0}\dot{z}_- + \omega_{-}^2z_{-} = \frac{F_{fluct, -}}{m}
\end{equation}
where $m_{1}=m_{2}=m$, $\gamma_{1}=\gamma_{2}=\gamma_{0}$ and $\langle F_{fluct, k}(t)F_{fluct, m}(t^{\prime})\rangle = 2m\gamma_{0}k_{B}T_{0}\delta(t - t^{\prime})\delta_{k,m}$ with $k, m = \{+, -\}$. Each mode will thermalise to the energy of the surrounding thermal bath i.e. $m\omega_{\pm}^{2}\langle z_{\pm}^{2}\rangle = k_{B}T_{0}$.

\par The motion of each particle will contain a fraction of the energy from each mode that is determined by the charge of that particle. By considering that energy is proportional to the variance of the displacement and using Eq.~\ref{eqn: EL2} we find the relations

\begin{equation}\label{eqn: E1}
    \frac{E_{1, +}}{E_{1}} = \frac{E_{2, -}}{E_{2}} = \frac{1}{r_{+}^2 + 1}
\end{equation}
\begin{equation}\label{eqn: E2}
    \frac{E_{2, +}}{E_{2}} = \frac{E_{1, -}}{E_{1}} = \frac{1}{r_{-}^{2} + 1}
\end{equation}
where $E_{i}$ represents the total energy of particle $i$ and $E_{i, k}$ represents the energy in particle $i$ coming from mode $k$. From this it can be seen that each particle will contain a total energy equal to the thermal bath. As the charge difference increases, $|r_{\pm}|\rightarrow \infty$ and $r_{\mp}\rightarrow 0$ and the particles no longer display normal modes. For $Q_{i}\gg Q_{j}$ we find $\omega_{-} = \omega_{i}$ and $\omega_{+} = \sqrt{3}\omega_{j}$ so the particle with large charge oscillates at its trap frequency and the particle with small charge is strongly affected by the electrostatic repulsion.

\par The radial motion of trapped nanoparticles will also couple to form normal modes, however, the coupling scales much more strongly with charge difference than the axial modes. For large charge differences both particles are almost completely unaffected by the other and oscillate close to their trap frequencies. The radial normal modes can be calculated in a similar manner to the axial normal modes~\cite{Wubbena2012}. 

\section{\label{sec:EM}Experimental Method}

\begin{figure*}
\centering
\includegraphics[scale=0.8]{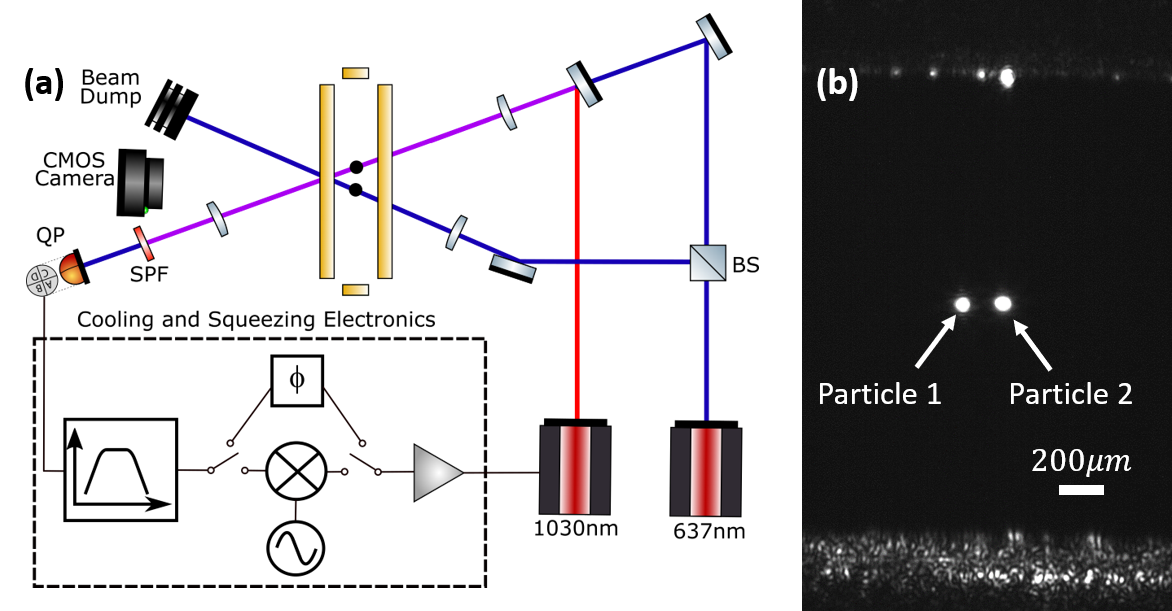}
\caption{(a) The experimental set-up for sympathetic cooling and squeezing. Two 387\,nm diameter, silica particles are trap simultaneously in a Paul trap. One arm of a 637\,nm laser is focused onto each particle individually. The power in each arm is balanced such that the scattering from each particle is approximately equal when measured on the CMOS camera. One of the arms is also focused onto a quadrant photodiode for real-time detection of the motion of one particle. A force is applied to the same particle to either cool or squeeze the normal modes by modulating the power of a 1030\,nm laser. The feedback signal is generated from the real-time measurement of the particle position. BS: Beam splitter, QP: Quadrant photodiode, SPF: Short-pass filter. (b) Two co-trapped particles in the Paul trap. The separation between these particles is $198\pm1$\,$\mu$m.}
\label{fig:ExpMeth}
\end{figure*}

\par In this experiment we use a linear Paul trap with four parallel rod electrodes for radial trapping and two "endcap" electrodes for axial trapping. The four parallel rods are held in place by two gold coated printed circuit boards which contain the electrical connections for the rods and have an endcap electrode etched into each~\cite{Bullier2020}. The parabolic coefficients were calculated using finite element modelling and found to be $r_0 = 1.1$\,mm, $z_0 = 3.5$\,mm, $\kappa = 0.071$ and $\eta = 0.82$. Typical trap parameters are $V_0 = 100 - 150$\,V, $U_0 = 20 - 50$\,V and $\omega_{rf} = 2\pi \times 8 - 12$\,kHz.

\par Silica nanoparticles, with charges of up to $6000e$, were loaded into the trap at $\sim 10^{-1}$\,mbar using the electrospray technique~\cite{Nagornykh2015, Bullier2020}. Two particles were either trapped simultaneously or, after trapping one particle, more nanoparticles were sprayed into the trapping region until a second was caught. Particles were monitored on a CMOS camera using scattered light from a $637$\,nm diode laser. Since both particles were illuminated by the 637\,nm laser (fig.~\ref{fig:ExpMeth}), timetraces of the particle motion could also be recorded on the CMOS camera at 1000 frames per second~\cite{Bullier2019, Bullier2020}. The timetraces were calibrated by moving the camera a fixed distance with a translation stage and recording the resulting displacement of the image. Both particles were recorded simultaneously in the same camera image so that the phase difference between the displacement of the particles was known. The camera acted as an out-of-loop detector for measuring the temperature when feedback cooling the particles.

\par Real-time detection of the particle motion was done using a quadrant photodiode. Individual arms of the 637\,nm beam illuminated each particle such that just the motion of one particle was measured on the quadrant photodiode. The signal from the quadrant photodiode was fed to a Red Pitaya FPGA to generate a feedback signal to either cool or squeeze the particle motion. The PyRPL software package was used to filter the position signal of the particle around the appropriate mode then either delay the signal (to cool the motion) or mix the signal with a sinusoidal wave at twice the central frequency of the mode (to squeeze the motion) followed by amplification. The feedback signal was then used to modulate the power of the 1030\,nm diode laser and create a force on the particle. Despite relatively high intensities of $2\times 10^{8}$\,Wm$^{-2}$ for the 1030\,nm laser, the trapping frequencies of particles were shifted by less than $2$\,\% due to the additional laser.

\section{\label{sec: NM}Particle Characterisation}

\begin{figure}
\centering
\includegraphics[scale=0.45]{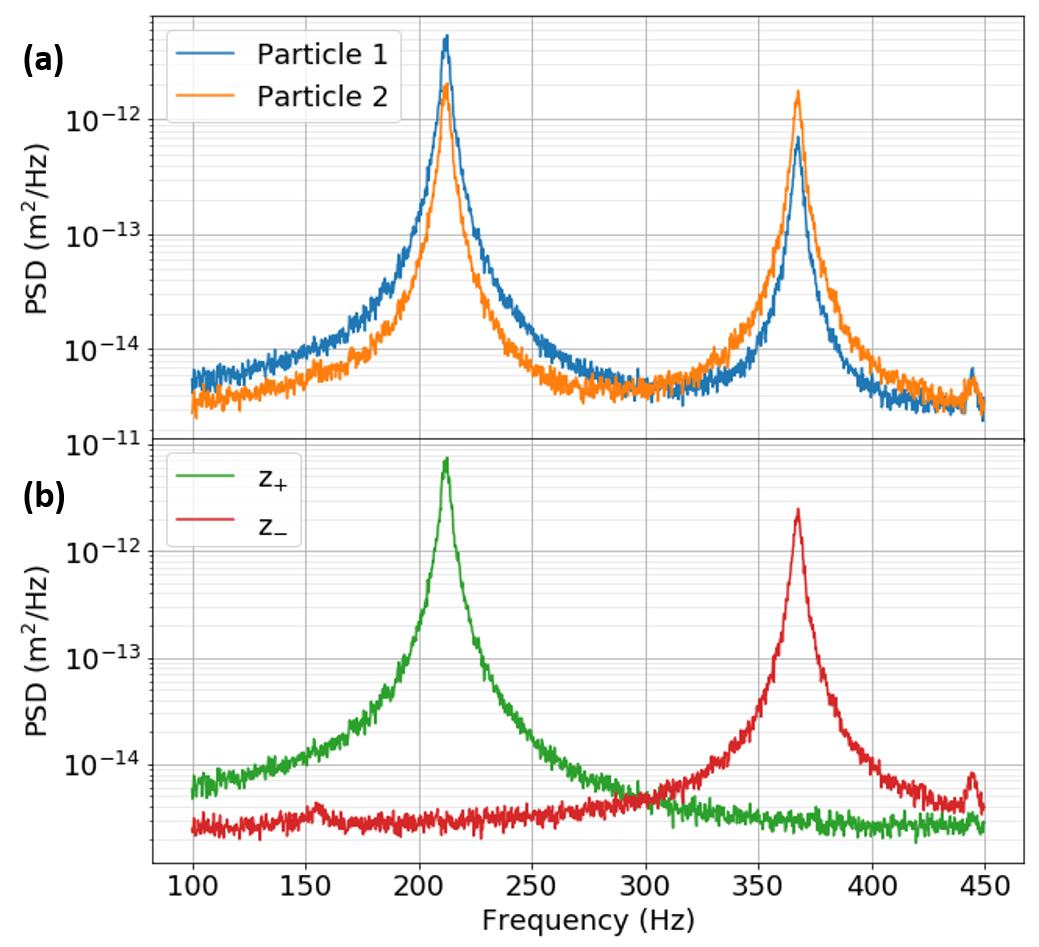}
\caption{a) PSDs of the motion of the two particles calculated from a timetrace taken on the CMOS camera. The imbalanced mode energies between each particle shows each particle has a different charge. b) PSDs of the normal modes of the two particle system. The normal modes are recovered from the motion of the individual particles by calculating the eigenvalues of the system. Neither mode has any component of the other mode present in the PSD therefore they are completely orthogonal and the masses of the two particles are approximately equal.}
\label{fig: NM}
\end{figure}

\par Fig.~\ref{fig:ExpMeth}b shows two particles trapped in the Paul trap. Using the CMOS camera, the equilibrium separation of the particles was measured to be $z_{sep}^{eq} = 198\pm1$\,$\mu$m. This is much larger than the expected amplitude of a single particle in thermal equilibrium with a frequency of $\omega_{0} = 2\pi\times 200$\,Hz ($\sqrt{\langle q^{2}\rangle} = 6.8$\,$\mu$m) so the Taylor expansion used in Eq.~\ref{eqn: EL} is valid. The power spectral density (PSD) of two coupled particles can be seen in fig.~\ref{fig: NM}a) taken at a pressure of $1.3\times10^{-2}$\,mbar. Unlike a single uncoupled particle which would display only one mode~\cite{Bullier2019, Bullier2020}, both particles display a mix of the normal modes of system showing they are axially coupled and it can be seen that they do not contain equal mode energy suggesting each particle has a different charge. The normal modes, $z_{\pm}$, are constructed from the measured timetraces using the linear transform described in Eq.~\ref{eqn: evecs}. To find the normal modes of the system the values of $r_{\pm}$ are varied until the PSD of the $z_{\pm}$ mode shows a minimal amount of the $z_{\mp}$ mode. Fig.~\ref{fig: NM}b) shows the PSDs of the normal modes for $r_{+} = 0.6$ and $r_{-} = -1.6$. Each normal mode is clearly seen with no component of the other mode suggesting the normal modes are orthogonal and the masses of each particle are approximately equal. The total energy in each particle was calculated by integrating the area under the PSDs of the individual particles and used to calculate their masses by assuming each particle is in thermal equilibrium with the surrounding gas. The particle radii are measured to be $r_{1} = 195\pm3$\,nm and $r_{2} = 192\pm3$\,nm assuming a density of $1850$\,kg/m$^{3}$. These values both agree with one another and agree with the nominal radius of $193.5$\,nm. Together, this is clear evidence that two single particles of approximately equal mass were trapped. Other pairs of trapped particle were measured to also have the mass of single particles with separations ranging from $150$\,$\mu$m to $200$\,$\mu$m.

\par The individual particle charges are different enough such that no coupling between the radial modes of each particle can be seen. This means the radial frequencies can be used to measure the individual charge-to-mass ratios of the two particles in the same manner as for a single trapped particle. By varying the frequency and voltage of the AC signal supplied to the rod electrodes whilst measuring the frequencies of the radial modes, charges of $Q_{1} = 2135\pm58$ \,$e$ and $Q_{2} = 906\pm15$\,$e$ were calculated using the mass values determined earlier. We can verify the charges by using them to calculate the theoretical values of $r_{\pm}$ and $E_{i,+}/E_{i}$ for the axial modes and comparing them to the measured values. We find $r_{-} = -1.60\pm0.03$ and $r_{+} = 0.61\pm0.04$ which are close to those used to construct the PSDs in fig.~\ref{fig: NM}b and that $\frac{E_{1,+}}{E_1} = \frac{E_{2,-}}{E_2} = 0.72\pm0.03$ which agree values  $\frac{E_{1,+}}{E_1} = 0.73\pm0.04$ and $\frac{E_{2,-}}{E_2} = 0.71\pm0.04$ measured from the PSDs. This shows that the radial modes have sufficiently small coupling to be treated as uncoupled.

\section{\label{sec: Cool}Sympathetic Cooling}

\begin{figure}
\centering
\includegraphics[scale=0.45]{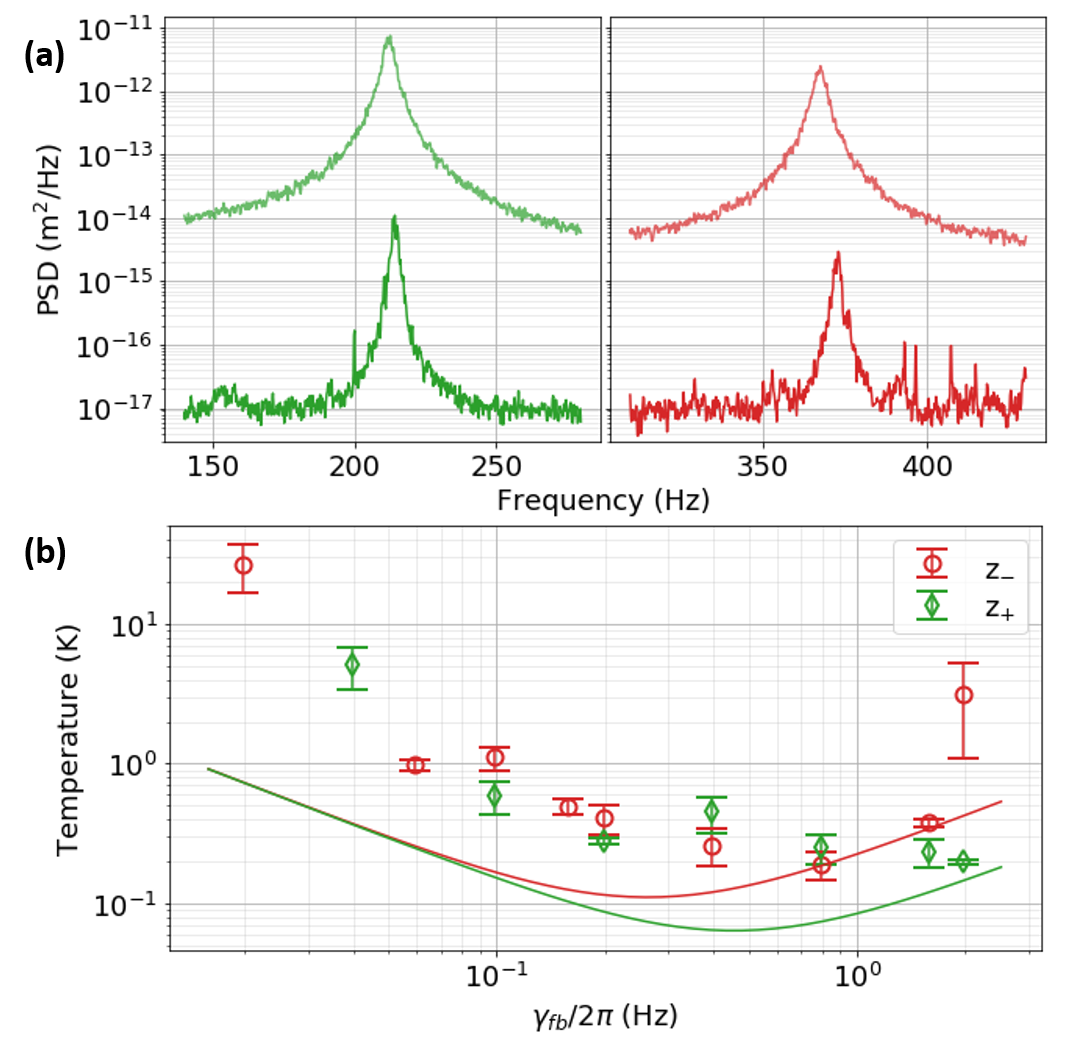}
\caption{a) Spectra for the cooled $z_{+}$ (left) and $z_{-}$ (right) modes at a pressure of $P=3.2\times 10^{-7}$\,mbar (dark lines) shown alongside the spectra of the modes with no cooling at $P=1.3\times 10^{-2}$\,mbar (light lines) where the modes are expected to be in thermal equilibrium with the surrounding gas. b) Temperature of the normal modes in the experiment (markers) and theory (lines) against feedback gain. The disagreement between theory and experiment suggests there is an additional source of heating for the normal modes.}
\label{fig: Cool}
\end{figure}

\par By modulating the power of a 1030\,nm laser focused onto just one of the particles (particle 1) the normal modes can be cooled. Through the Coulomb interaction, the same modes of the other particle are also cooled. If a force proportional to the velocity of the $z_{+}$ is applied to particle 1, the resulting EoM for the two normal modes are
\begin{equation}
    \ddot{z}_{+} + (\gamma_{0}+\gamma_{fb})\dot{z}_+ + \gamma_{fb}\delta\dot{z}_{+} + \omega_{+}^2z_{+} = \frac{F_{fluct, +}}{m}
\end{equation}
\begin{equation}\label{eqn: modz-}
    \ddot{z}_{-} + \gamma_{0}\dot{z}_- - \gamma_{fb}\dot{z}_{+} - \gamma_{fb}\delta\dot{z}_{+} + \omega_{-}^2z_{-} = \frac{F_{fluct, -}}{m}
\end{equation}
where $\gamma_{fb}$ is the feedback gain and $ \delta\dot{z}_{+}$ is imprecision noise in the detection. Provided $\omega_{-}-\omega_{+}\gg\gamma_{0}+\gamma_{fb}$, the $\gamma_{fb}\dot{z}_{+}$ term in Eq. \ref{eqn: modz-} will have a negligible effect on the $z_{-}$ mode and can be ignored. Additionally, in practice the feedback signal is bandpass filtered such that the $\gamma_{fb}\delta\dot{z}_{+}$ term does not affect the $z_{-}$ mode. Thus, the $z_{+}$ mode is cooled whilst the $z_{-}$ mode remains unaffected.Similar equations can be found for the two modes (by interchaging all $\pm$ subscripts) when cooling the $z_{-}$ mode. Since the particles are the same mass, the temperature compression ratio for a mode should be equal in each particle~\cite{Wubbena2012}. The cooled mode will have a temperature given by~\cite{Tebbenjohans2018, Dania2020, Kamba2021, Penny2021} 

\begin{equation}\label{eqn: vdTCoM}
    T_{+} = T_{0, +}\frac{\gamma_{0}}{\gamma_{0} + \gamma_{fb}} + \frac{1}{2}\frac{m\omega_{+}^{2}}{k_{B}}\frac{\gamma_{fb}^{2}}{\gamma_{0} + \gamma_{fb}}S_{nn, +}
\end{equation}
where $T_{0,+}$ is the initial temperature of the CoM mode and $S_{nn, +} =  \int_{-\infty}^{\infty}\delta z_{+}(t)\delta z_{+}(0)e^{i\omega t}dt$ is the spectral density of the imprecision noise and is assumed to be white and Gaussian over the linewidth of the oscillator mode. A similar expression is found for the $z_{-}$ mode. By increasing the feedback gain a minimum temperature will be found that is dependent on the frequency of the mode.

\par In the experiment, the feedback was implemented by applying a bandpass filter  to the position signal of the particle to remove noise and other modes from the feedback signal then amplifying and adding a $\pi/2$ phase shift to estimate the current velocity of the particle~\cite{Dania2020, Penny2021}. Fig.~\ref{fig: Cool}a) shows the spectra of the two modes cooled to minimum temperatures of $T_{+} = 200\pm10$\,mK and $T_{-} = 190\pm40$\,mK (calculated from the area under the PSDs) at a pressure of $P = 3.2\times10^{-7}$\,mbar and spectra of the modes with no cooling ($T_{+} = T_{-} =293$\,K) at $P = 1.3\times10^{-2}$\,mbar. As expected, the temperature compression ratio of a mode was found to be the same for each particle. The final temperatures reached are approximately twice as high as the expected temperatures from measurements of detection noise and pressure. By comparing the measured mode temperature as a function of feedback gain to the theoretical prediction (fig.~\ref{fig: Cool}b) we see that the temperatures are higher at low gain than expected. This suggests additional heating of the particle motion which calculations show could be partly due to voltage noise in the electronics. The voltage noise is increased compared to a single particle since the particles are pushed off-centre by their Coulomb repulsion. Alternative trap geometries could be used to reduce this effect. At a pressure of $10^{-1}$\,mbar, where the particle mass was measured, this level of voltage noise would have a negligible effect on the motion of the particle. 

\par To decrease the temperature further, the detection noise could be improved or the pressure could be reduced further. The additional force noise would have to be removed before reducing the pressure since white, Gaussian noise sources increase the temperature of the oscillator scaling with $1/\gamma_{0}$ whereas velocity damping only scales with $\sqrt{\gamma_{0}}$. Here, we measure a detection noise of $S_{nn} = 3\times10^{-15}$\,m$^{2}$Hz$^{-1}$ which limits the final temperature at a given pressure. Recent experiments with single particles in Paul traps have demonstrated detection noise as low as $2.9\times10^{-24}$\,m$^{2}$Hz$^{-1}$~\cite{Dania2022}. Using a similar detection technique and a numerical aperture (NA) limited by the trap geometry (NA = 0.5) would allow us to reach a minimum occupancy of $\bar{n} \sim 0.5$~\cite{Cerchiari2021} with similar additional optical losses to other experiments~\cite{Magrini2021}. In order to remain in the underdamped regime where the theory is valid, the oscillator frequency would have to be increased and the background gas damping would have to be reduced. An oscillator at a frequency of $\sim 500$\,Hz and a background pressure of $\sim 10^{-11}$\,mbar would be sufficient provided other noise sources are kept negligible. 

\section{\label{sec: Squeeze}Sympathetic Squeezing}

\begin{figure}
\centering
\includegraphics[scale=0.42]{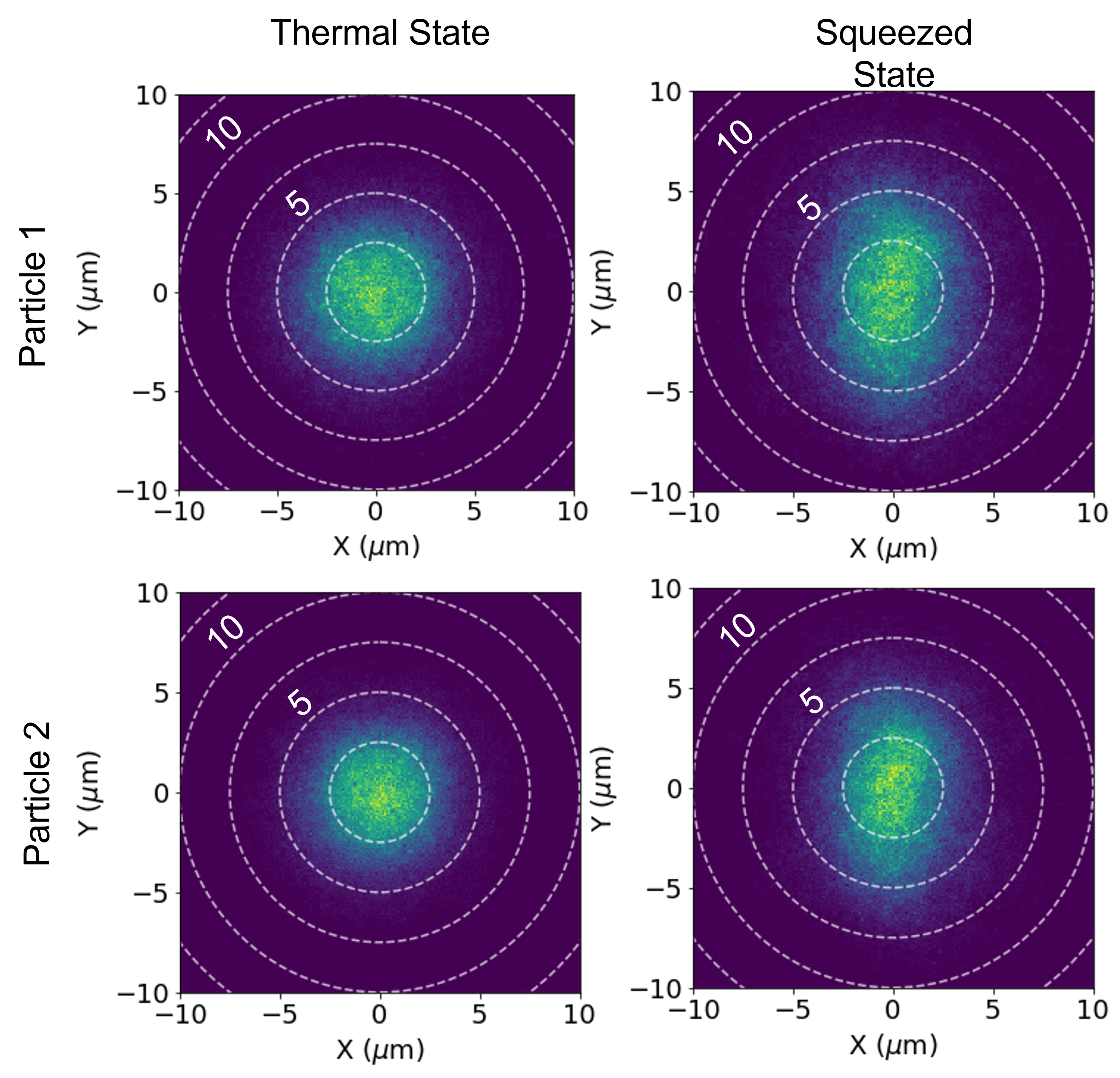}
\caption{Phase space diagrams showing a thermal and squeezed state for the $z_{+}$ mode of both particles at a pressure of $P=1.2\times10^{-2}$\,mbar. Both particles display squeezed states despite interaction with only particle 1. The particles used here have approximately equal charges therefore the mode energies should be equal as they appear here.}
\label{fig: Squeeze}
\end{figure}

\par Squeezing the motion of a mechanical oscillator has been achieved with several methods including parametric modulation~\cite{Rugar1991, Briant2003, Szorkovsky2011, Szorkovszky2012, Szorkovszky2013, Pontin2014, Poot2014, Pontin2016}, non-adiabatic shifts~\cite{Asjad2014, Rashid2016, Setter2019} and back-action evading measurements~\cite{Braginsky1980, Clerk2008, Hertzberg2010, Vanner2013}. Parametric modulation is usually implemented by modulating the oscillator spring constant at twice the mode resonance frequency. In the Paul trap this would drive both particles. In order to show a sympathetic operation, we use a measurement based scheme to parametrically drive only one particle. This scheme also avoids acting on either particle with a linear drive that would appear from modulating the trap potential~\cite{Gehm1998, Gorman2014}. A measurement of the particle position was filtered around the appropriate mode using a Lorentzian bandpass filter with a bandwidth of $152$\,Hz then mixed with a sinusoidal signal at twice the resonance frequency of that mode to produce a signal to modulate the 1030\,nm laser. An additional delay was added to the signal to minimise errors to the phase response of the filter. The laser imparted a force $F\propto z_{\pm}\text{sin}(2\omega_{\pm}t)$ onto particle 1. If we were, for example, squeezing the $z_{+}$ mode this results in the following EoM for the normal modes
\begin{equation}
        \ddot{z}_{+} + \gamma_{0}\dot{z}_+ + \omega_{+}^2z_{+} + (z_{+}+\delta z_{+})G\text{sin}(2\omega_{+}t) = \frac{F_{fluct, +}}{m}
\end{equation}
\begin{equation}
    \ddot{z}_{-} + \gamma_{0}\dot{z}_- + \omega_{-}^2z_{-} + (z_{+}+\delta z_{+})G\text{sin}(2\omega_{+}t) = \frac{F_{fluct, -}}{m}
\end{equation}
where $G$ is a gain applied to the signal. Similar to the case for sympathetic cooling, the $z_{-}$ mode remains unaffected by the applied force provided $\omega_{-} - \omega_{+} \gg \gamma_{0}$. In this instance, $\delta z_{+} \ll z_{+}$ and therefore has a negligible impact on the dynamics of the system. Thus, the $z_{+}$ mode experiences a parametric drive whilst the $z_{-}$ mode is unaffected. In the case of squeezing with a parametric drive, the variance of the X and Y quadratures is given by~\cite{Briant2003}

\begin{equation}
    \langle X^{2} \rangle = \frac{\langle x^{2}\rangle}{1-g}
\end{equation}

\begin{equation}
    \langle Y^{2} \rangle = \frac{\langle x^{2}\rangle}{1+g}
\end{equation}
where $g = G\times \omega_{0}/2\gamma_{0}$. It can be seen that this limits the maximum achievable squeezing to -3 dB before the onset of parametric instability in the X quadrature~\cite{Collett1984, walls2012}.

\par Fig.~\ref{fig: Squeeze} shows phase space plots of the quadratures of the $z_{+}$ mode of both particles with and without a parametric drive applied to particle 1 at a pressure of $P=1.2\times10^{-2}$\,mbar. Squeezing was performed at a relatively high pressure to reduce the measurement time required to accurately sample the squeezed thermal state. The quadratures were extracted from the displacement data by applying a demodulation followed by a $50$\,Hz low-pass filter to the $z_{+}$ mode and sweeping the demodulation frequency until maximal squeezing was observed. A clear signature of squeezing can be seen in the elongations of the phase space distributions in both particles. Anomalous heating of the mode was also seen as the gain of the squeezing operation was increased. Simulations based on the experimental parameters, implemented with the leapfrog algorithm, suggest that suggest this is due to using a tightly focused beam to generate the force for squeezing the particle. Since the beam waist is comparable to the amplitude of particle motion the particle feels an additional position dependent component to the feedback force. Taking the heating into account, we were able to measure $-1.70\pm 0.05$\,dB and $-1.69\pm 0.05$\,dB of squeezing in the $z_{+}$ mode of particle 1 and 2 respectively. The squeezing gain was limited by the non-linear effect of the tight beam generating a spiralling in the phase space plots \cite{Setter2019}. Both the $z_{+}$ and $z_{-}$ mode could be squeezed using this method.

\par Transferal of the thermal squeezed state between the particles shows that non-thermal states can also be transmitted via the Coulomb force. For an array of coupled nanoparticles this property could be used to control and readout the state of the array from only a single nanoparticle. For large arrays, this could be technically easier than reading out the state of every single nanoparticle. 

\section{\label{sec:conc}Conclusions}

\par We have demonstrated and characterised coupling between two charged nanoparticles that produces two orthogonal normal modes of motion. Both sympathetic cooling and squeezing of the motion of one particle were shown through interaction with another coupled particle. This demonstrates thermal and non-thermal states can be transferred via the Coulomb interaction. Such techniques could be extended to a mixed species system with particles of similar masses or to an array of many nanoparticles. As a result, this work represents an important tool in implementing the creation of macropscopic quantum superpositions~\cite{Scala2013, Yin2013, Yin2015, WanQSDrop2016, Sougato2017} via coupling of the normal modes of a co-trapped silica-nanodiamond pair to the internal quantum state of a nitrogen vacancy centre in the nanodiamond. Such a technique can be applied to experiments which have previously been limited by internal heating of samples~\cite{Rahman2016}.

\par Our results are a first step towards schemes that utilize the coupling between levitated particles. Large arrays of coupled particles would have a range of motional frequencies that scale with the number of particles. This would be ideal for sensing over a wide range of frequencies simultaneously such as in ultra-light dark matter searches~\cite{Carney2021}. Using sympathetic techniques would allow control and read-out of the entire array from a single particle simplifying the experimental procedure compared to an uncoupled array. 


\section*{Acknowledgments}

The authors would like to acknowledge useful discussions with Animesh Datta. The authors acknowledge funding from the EPSRC Grant No. EP/N031105/1  and  the  H2020-EU.1.2.1  TEQ  project Grant agreement ID: 766900.


\bibliography{bibliography}

\begin{thebibliography}{10}

\bibitem{Monteiro2020}
F.~Monteiro, G.~Afek, D.~Carney, G.~Krnjaic, J.~Wang, and D.~C. Moore, ``Search
  for composite dark matter with optically levitated sensors,'' {\em Phys. Rev.
  Lett.}, vol.~125, p.~181102, Oct 2020.

\bibitem{Moore2021}
D.~C. Moore and A.~A. Geraci, ``Searching for new physics using optically
  levitated sensors,'' {\em Quantum Science and Technology}, vol.~6, p.~014008,
  jan 2021.

\bibitem{Carney2021}
D.~Carney, G.~Krnjaic, D.~C. Moore, C.~A. Regal, G.~Afek, S.~Bhave,
  B.~Brubaker, T.~Corbitt, J.~Cripe, N.~Crisosto, A.~Geraci, S.~Ghosh, J.~G.~E.
  Harris, A.~Hook, E.~W. Kolb, J.~Kunjummen, R.~F. Lang, T.~Li, T.~Lin, Z.~Liu,
  J.~Lykken, L.~Magrini, J.~Manley, N.~Matsumoto, A.~Monte, F.~Monteiro,
  T.~Purdy, C.~J. Riedel, R.~Singh, S.~Singh, K.~Sinha, J.~M. Taylor, J.~Qin,
  D.~J. Wilson, and Y.~Zhao, ``Mechanical quantum sensing in the search for
  dark matter,'' {\em Quantum Science and Technology}, vol.~6, p.~024002, jan
  2021.

\bibitem{RomeroIsart2010}
O.~Romero-Isart, M.~L. Juan, R.~Quidant, and J.~I. Cirac, ``Toward quantum
  superposition of living organisms,'' {\em New Journal of Physics}, vol.~12,
  p.~033015, mar 2010.

\bibitem{BatemanWMInter2014}
J.~Bateman, S.~Nimmrichter, and K.~e.~a. Hornberger, ``Near-field
  interferometry of a free-falling nanoparticle from a point-like source,''
  {\em Nat. Comm.}, vol.~5, no.~4788, 2014.

\bibitem{Bahrami2014}
M.~Bahrami, M.~Paternostro, A.~Bassi, and H.~Ulbricht, ``Proposal for a
  noninterferometric test of collapse models in optomechanical systems,'' {\em
  Phys. Rev. Lett.}, vol.~112, p.~210404, May 2014.

\bibitem{Weiss2021}
T.~Weiss, M.~Roda-Llordes, E.~Torrontegui, M.~Aspelmeyer, and O.~Romero-Isart,
  ``Large quantum delocalization of a levitated nanoparticle using optimal
  control: Applications for force sensing and entangling via weak forces,''
  {\em Phys. Rev. Lett.}, vol.~127, p.~023601, Jul 2021.

\bibitem{Cosco2021}
F.~Cosco, J.~S. Pedernales, and M.~B. Plenio, ``Enhanced force sensitivity and
  entanglement in periodically driven optomechanics,'' {\em Phys. Rev. A},
  vol.~103, p.~L061501, Jun 2021.

\bibitem{GeraciZepto2016}
G.~Ranjit, M.~Cunningham, K.~Casey, and A.~A. Geraci, ``Zeptonewton force
  sensing with nanospheres in an optical lattice,'' {\em Phys. Rev. A},
  vol.~93, p.~053801, May 2016.

\bibitem{Hebestreit2018_2}
E.~Hebestreit, M.~Frimmer, R.~Reimann, and L.~Novotny, ``Sensing static forces
  with free-falling nanoparticles,'' {\em Phys. Rev. Lett.}, vol.~121,
  p.~063602, Aug 2018.

\bibitem{Kawasaki2020}
A.~Kawasaki, A.~Fieguth, N.~Priel, C.~P. Blakemore, D.~Martin, and G.~Gratta,
  ``High sensitivity, levitated microsphere apparatus for short-distance force
  measurements,'' {\em Review of Scientific Instruments}, vol.~91, no.~8,
  p.~083201, 2020.

\bibitem{Arita2018}
Y.~Arita, E.~M. Wright, and K.~Dholakia, ``Optical binding of two cooled
  micro-gyroscopes levitated in vacuum,'' {\em Optica}, vol.~5, pp.~910--917,
  Aug 2018.

\bibitem{Slezak2019}
B.~R. Slezak and B.~D'Urso, ``A microsphere molecule: The interaction of two
  charged microspheres in a magneto-gravitational trap,'' {\em Applied Physics
  Letters}, vol.~114, no.~24, p.~244102, 2019.

\bibitem{Zhao2012}
R.~Zhao, A.~Manjavacas, F.~J. Garc\'{\i}a~de Abajo, and J.~B. Pendry,
  ``Rotational quantum friction,'' {\em Phys. Rev. Lett.}, vol.~109, p.~123604,
  Sep 2012.

\bibitem{Sougato2017}
S.~Bose, A.~Mazumdar, G.~W. Morley, H.~Ulbricht, M.~Toro\ifmmode~\check{s}\else
  \v{s}\fi{}, M.~Paternostro, A.~A. Geraci, P.~F. Barker, M.~S. Kim, and
  G.~Milburn, ``Spin entanglement witness for quantum gravity,'' {\em Phys.
  Rev. Lett.}, vol.~119, p.~240401, Dec 2017.

\bibitem{Datta2021}
A.~Datta and H.~Miao, ``Signatures of the quantum nature of gravity in the
  differential motion of two masses,'' {\em Quantum Science and Technology},
  vol.~6, p.~045014, aug 2021.

\bibitem{King1998}
B.~E. King, C.~S. Wood, C.~J. Myatt, Q.~A. Turchette, D.~Leibfried, W.~M.
  Itano, C.~Monroe, and D.~J. Wineland, ``Cooling the collective motion of
  trapped ions to initialize a quantum register,'' {\em Phys. Rev. Lett.},
  vol.~81, pp.~1525--1528, Aug 1998.

\bibitem{Wineland1978}
D.~J. Wineland, R.~E. Drullinger, and F.~L. Walls, ``Radiation-pressure cooling
  of bound resonant absorbers,'' {\em Phys. Rev. Lett.}, vol.~40,
  pp.~1639--1642, Jun 1978.

\bibitem{Drullinger1980}
R.~E. Drullinger, D.~J. Wineland, and J.~C. Bergquist, ``High- resolution
  optical spectra of laser cooled ions,'' {\em Appl. Phys.}, vol.~22,
  pp.~365--368, 1980.

\bibitem{Larson1986}
D.~J. Larson, J.~C. Bergquist, J.~J. Bollinger, W.~M. Itano, and D.~J.
  Wineland, ``Sympathetic cooling of trapped ions: A laser-cooled two-species
  nonneutral ion plasma,'' {\em Phys. Rev. Lett.}, vol.~57, pp.~70--73, Jul
  1986.

\bibitem{Baba1996}
T.~Baba and I.~Waki, ``Cooling and mass-analysis of molecules using
  laser-cooled atoms,'' {\em Japanese Journal of Applied Physics}, vol.~35,
  pp.~L1134--L1137, sep 1996.

\bibitem{Molhave2000}
K.~M\o{}lhave and M.~Drewsen, ``Formation of translationally cold
  ${\mathrm{mgh}}^{+}$ and ${\mathrm{mgd}}^{+}$ molecules in an ion trap,''
  {\em Phys. Rev. A}, vol.~62, p.~011401(R), Jun 2000.

\bibitem{Millen2014}
J.~Millen, T.~Deesuwan, P.~F. Barker, and J.~Anders, ``Nanoscale temperature
  measurements using non-equilibrium brownian dynamics of a levitated
  nanosphere,'' {\em Nature Nanotechnology}, vol.~9, pp.~425--429, 2014.

\bibitem{Rahman2016}
A.~T.~M. Anishur~Rahman, A.~Frangeskou, K.~M., S.~Bose, G.~W. Morley, and P.~F.
  Barker, ``Burning and graphitization of optically levitated nanodiamonds in
  vacuum,'' {\em Sci Rep}, vol.~6, p.~21633, 2016.

\bibitem{kimsqueezed}
M.~S. Kim, F.~A.~M. de~Oliveira, and P.~L. Knight, ``Properties of squeezed
  number states and squeezed thermal states,'' {\em Phys. Rev. A}, vol.~40,
  pp.~2494--2503, Sep 1989.

\bibitem{science_squeezing}
E.~E. Wollman, C.~U. Lei, A.~J. Weinstein, J.~Suh, A.~Kronwald, F.~Marquardt,
  A.~A. Clerk, and K.~C. Schwab, ``Quantum squeezing of motion in a mechanical
  resonator,'' vol.~349, pp.~952--955, Aug. 2015.

\bibitem{Pontin2016}
A.~Pontin, M.~Bonaldi, A.~Borrielli, L.~Marconi, F.~Marino, G.~Pandraud, G.~A.
  Prodi, P.~M. Sarro, E.~Serra, and F.~Marin, ``Dynamical two-mode squeezing of
  thermal fluctuations in a cavity optomechanical system,'' {\em Phys. Rev.
  Lett.}, vol.~116, p.~103601, Mar 2016.

\bibitem{Poot2014}
M.~Poot, K.~Y. Fong, and H.~X. Tang, ``Classical non-gaussian state preparation
  through squeezing in an optoelectromechanical resonator,'' {\em Phys. Rev.
  A}, vol.~90, p.~063809, Dec 2014.

\bibitem{caves_RevModPhys.52.341}
C.~M. Caves, K.~S. Thorne, R.~W.~P. Drever, V.~D. Sandberg, and M.~Zimmermann,
  ``On the measurement of a weak classical force coupled to a
  quantum-mechanical oscillator. i. issues of principle,'' {\em Rev. Mod.
  Phys.}, vol.~52, pp.~341--392, Apr 1980.

\bibitem{parametric_amplification}
Q.~P. Unterreithmeier, E.~M. Weig, and J.~P. Kotthaus, ``Universal transduction
  scheme for nanomechanical systems based on dielectric forces,'' vol.~458,
  pp.~1001--1004, Apr. 2009.

\bibitem{Li2011}
T.~Li, S.~Kheifets, and M.~Raizen, ``Millikelvin cooling of an optically
  trapped microsphere in vacuum,'' {\em Nature Phys}, vol.~7, pp.~527--530,
  2011.

\bibitem{Tebbenjohans2018}
F.~Tebbenjohanns, M.~Frimmer, A.~Militaru, V.~Jain, and L.~Novotny, ``Cold
  damping of an optically levitated nanoparticle to microkelvin temperatures,''
  {\em Phys. Rev. Lett.}, vol.~122, p.~223601, Jun 2019.

\bibitem{Slezak2018}
B.~R. Slezak, C.~W. Lewandowski, J.-F. Hsu, and B.~D'Urso, ``Cooling the motion
  of a silica microsphere in a magneto-gravitational trap in ultra-high
  vacuum,'' {\em New Journal of Physics}, vol.~20, p.~063028, jun 2018.

\bibitem{Iwasaki2019}
M.~Iwasaki, T.~Yotsuya, T.~Naruki, Y.~Matsuda, M.~Yoneda, and K.~Aikawa,
  ``Electric feedback cooling of single charged nanoparticles in an optical
  trap,'' {\em Phys. Rev. A}, vol.~99, p.~051401(R), May 2019.

\bibitem{Dania2020}
L.~Dania, D.~S. Bykov, M.~Knoll, P.~Mestres, and T.~E. Northup, ``Optical and
  electrical feedback cooling of a silica nanoparticle levitated in a paul
  trap,'' {\em Phys. Rev. Research}, vol.~3, p.~013018, Jan 2021.

\bibitem{Tebbenjohans2020}
F.~Tebbenjohanns, M.~Frimmer, V.~Jain, D.~Windey, and L.~Novotny, ``Motional
  sideband asymmetry of a nanoparticle optically levitated in free space,''
  {\em Phys. Rev. Lett.}, vol.~124, p.~013603, Jan 2020.

\bibitem{Kamba2021}
M.~Kamba, H.~Kiuchi, T.~Yotsuya, and K.~Aikawa, ``Recoil-limited feedback
  cooling of single nanoparticles near the ground state in an optical
  lattice,'' {\em Phys. Rev. A}, vol.~103, p.~L051701, May 2021.

\bibitem{Tebbenjohans2021}
F.~Tebbenjohanns, M.~Mattana, M.~Rossi, M.~Frimmer, and L.~Novotny, ``Quantum
  control of a nanoparticle optically levitated in cryogenic free space,'' {\em
  Nature}, vol.~595, pp.~378--382, 2021.

\bibitem{Penny2021}
T.~W. Penny, A.~Pontin, and P.~F. Barker, ``Performance and limits of feedback
  cooling methods for levitated oscillators: A direct comparison,'' {\em Phys.
  Rev. A}, vol.~104, p.~023502, Aug 2021.

\bibitem{Rugar1991}
D.~Rugar and P.~Gr\"utter, ``Mechanical parametric amplification and
  thermomechanical noise squeezing,'' {\em Phys. Rev. Lett.}, vol.~67,
  pp.~699--702, Aug 1991.

\bibitem{Briant2003}
T.~Briant, P.~Cohadon, M.~Pinard, and A.~Heidmann, ``Optical phase-space
  reconstruction of mirror motion at the attometer level,'' {\em Eur. Phys. J.
  D.}, vol.~22, pp.~131--140, Jan 2003.

\bibitem{Szorkovsky2011}
A.~Szorkovszky, A.~C. Doherty, G.~I. Harris, and W.~P. Bowen, ``Mechanical
  squeezing via parametric amplification and weak measurement,'' {\em Phys.
  Rev. Lett.}, vol.~107, p.~213603, Nov 2011.

\bibitem{Szorkovszky2012}
A.~Szorkovszky, A.~C. Doherty, G.~I. Harris, and W.~P. Bowen, ``Position
  estimation of a parametrically driven optomechanical system,'' {\em New
  Journal of Physics}, vol.~14, p.~095026, sep 2012.

\bibitem{Szorkovszky2013}
A.~Szorkovszky, G.~A. Brawley, A.~C. Doherty, and W.~P. Bowen, ``Strong
  thermomechanical squeezing via weak measurement,'' {\em Phys. Rev. Lett.},
  vol.~110, p.~184301, May 2013.

\bibitem{Pontin2014}
A.~Pontin, M.~Bonaldi, A.~Borrielli, F.~S. Cataliotti, F.~Marino, G.~A. Prodi,
  E.~Serra, and F.~Marin, ``Squeezing a thermal mechanical oscillator by
  stabilized parametric effect on the optical spring,'' {\em Phys. Rev. Lett.},
  vol.~112, p.~023601, Jan 2014.

\bibitem{Morigi2001}
G.~Morigi and H.~Walther, ``Two-species coulomb chains for quantum
  information,'' {\em Eur. Phys. J. D}, vol.~13, pp.~261--269, 2001.

\bibitem{Wubbena2012}
J.~B. W\"ubbena, S.~Amairi, O.~Mandel, and P.~O. Schmidt, ``Sympathetic cooling
  of mixed-species two-ion crystals for precision spectroscopy,'' {\em Phys.
  Rev. A}, vol.~85, p.~043412, Apr 2012.

\bibitem{Bullier2020}
N.~P. Bullier, A.~Pontin, and P.~F. Barker, ``Characterisation of a charged
  particle levitated nano-oscillator,'' {\em Journal of Physics D: Applied
  Physics}, vol.~53, p.~175302, feb 2020.

\bibitem{Nagornykh2015}
P.~Nagornykh, J.~E. Coppock, and B.~E. Kane, ``Cooling of levitated graphene
  nanoplatelets in high vacuum,'' {\em Applied Physics Letters}, vol.~106,
  no.~24, p.~244102, 2015.

\bibitem{Bullier2019}
N.~P. Bullier, A.~Pontin, and P.~F. Barker, ``Super-resolution imaging of a low
  frequency levitated oscillator,'' {\em Review of Scientific Instruments},
  vol.~90, no.~9, p.~093201, 2019.

\bibitem{Dania2022}
L.~Dania, K.~Heidegger, D.~S. Bykov, G.~Cerchiari, G.~Araneda, and T.~E.
  Northup, ``Position measurement of a levitated nanoparticle via interference
  with its mirror image,'' {\em Phys. Rev. Lett.}, vol.~129, p.~013601, Jun
  2022.

\bibitem{Cerchiari2021}
G.~Cerchiari, L.~Dania, D.~S. Bykov, R.~Blatt, and T.~E. Northup, ``Position
  measurement of a dipolar scatterer via self-homodyne detection,'' {\em Phys.
  Rev. A}, vol.~104, p.~053523, Nov 2021.

\bibitem{Magrini2021}
L.~Magrini, P.~Rosenzweig, C.~Bach, A.~Deutschmann-Olek, S.~G. Hofer, S.~Hong,
  N.~Kiesel, A.~Kugi, and M.~Aspelmeyer, ``Real-time optimal quantum control of
  mechanical motion at room temperature,'' {\em Nature}, vol.~595,
  pp.~373--377, 2021.

\bibitem{Asjad2014}
M.~Asjad, G.~S. Agarwal, M.~S. Kim, P.~Tombesi, G.~Di~Giuseppe, and D.~Vitali,
  ``Robust stationary mechanical squeezing in a kicked quadratic optomechanical
  system,'' {\em Phys. Rev. A}, vol.~89, p.~023849, Feb 2014.

\bibitem{Rashid2016}
M.~Rashid, T.~Tufarelli, J.~Bateman, J.~Vovrosh, D.~Hempston, M.~S. Kim, and
  H.~Ulbricht, ``Experimental realization of a thermal squeezed state of
  levitated optomechanics,'' {\em Phys. Rev. Lett.}, vol.~117, p.~273601, Dec
  2016.

\bibitem{Setter2019}
A.~Setter, J.~Vovrosh, and H.~Ulbricht, ``Characterization of non-linearities
  through mechanical squeezing in levitated optomechanics,'' {\em Applied
  Physics Letters}, vol.~115, no.~15, p.~153106, 2019.

\bibitem{Braginsky1980}
V.~B. Braginsky, Y.~I. Vorontsovand, and K.~S. Thorne, ``Quantum nondemolition
  measurements,'' {\em Science}, vol.~209, pp.~547--557, 1980.

\bibitem{Clerk2008}
A.~A. Clerk, F.~Marquardt, and K.~Jacobs, ``Back-action evasion and squeezing
  of a mechanical resonator using a cavity detector,'' {\em New Journal of
  Physics}, vol.~10, p.~095010, sep 2008.

\bibitem{Hertzberg2010}
J.~B. Hertzberg, T.~Rocheleau, T.~Ndukum, M.~Savva, A.~A. Clerk, and K.~C.
  Schwab, ``Back-action-evading measurements of nanomechanical motion,'' {\em
  Nature Physics}, vol.~6, pp.~213--217, 2010.

\bibitem{Vanner2013}
M.~R. Vanner, J.~Hofer, G.~D. Cole, and M.~Aspelmeyer, ``Cooling-by-measurement
  and mechanical state tomography via pulsed optomechanics,'' {\em Nat Commun},
  vol.~4, p.~2295, 2013.

\bibitem{Gehm1998}
M.~E. Gehm, K.~M. O'Hara, T.~A. Savard, and J.~E. Thomas, ``Dynamics of
  noise-induced heating in atom traps,'' {\em Phys. Rev. A}, vol.~58,
  pp.~3914--3921, Nov 1998.

\bibitem{Gorman2014}
D.~J. Gorman, P.~Schindler, S.~Selvarajan, N.~Daniilidis, and H.~H\"affner,
  ``Two-mode coupling in a single-ion oscillator via parametric resonance,''
  {\em Phys. Rev. A}, vol.~89, p.~062332, Jun 2014.

\bibitem{Collett1984}
M.~J. Collett and C.~W. Gardiner, ``Squeezing of intracavity and traveling-wave
  light fields produced in parametric amplification,'' {\em Phys. Rev. A},
  vol.~30, pp.~1386--1391, Sep 1984.

\bibitem{walls2012}
D.~Walls and G.~Milburn, {\em Quantum Optics}.
\newblock Springer Study Edition, Springer Berlin Heidelberg, 2012.

\bibitem{Scala2013}
M.~Scala, M.~S. Kim, G.~W. Morley, P.~F. Barker, and S.~Bose, ``Matter-wave
  interferometry of a levitated thermal nano-oscillator induced and probed by a
  spin,'' {\em Phys. Rev. Lett.}, vol.~111, p.~180403, Oct 2013.

\bibitem{Yin2013}
Z.-q. Yin, T.~Li, X.~Zhang, and L.~M. Duan, ``Large quantum superpositions of a
  levitated nanodiamond through spin-optomechanical coupling,'' {\em Phys. Rev.
  A}, vol.~88, p.~033614, Sep 2013.

\bibitem{Yin2015}
Z.~Yin, N.~Zhao, and T.~Li, ``Hybrid opto-mechanical systems with
  nitrogen-vacancy centers,'' {\em Sci. China Phys. Mech. Astron.}, vol.~58,
  pp.~1--12, 2015.

\bibitem{WanQSDrop2016}
C.~Wan, M.~Scala, G.~W. Morley, {\relax ATM}.~A. Rahman, H.~Ulbricht,
  J.~Bateman, P.~F. Barker, S.~Bose, and M.~S. Kim, ``Free nano-object ramsey
  interferometry for large quantum superpositions,'' {\em Phys. Rev. Lett.},
  vol.~117, p.~143003, Sep 2016.

\end{thebibliography}

\end{document}